\documentclass[11pt,tightenlines,aps,prd,superscriptaddress,
nofootinbib,nobibnotes,longbibliography,notitlepage]{revtex4-1}
\pdfoutput=1

\usepackage{graphicx}
\usepackage{amsfonts}
\usepackage{amssymb}
\usepackage{amsbsy}
\usepackage{amsmath}
\usepackage{mathtools}
\usepackage{latexsym}
\usepackage{bm}
\usepackage{color}
\usepackage{hyperref}
\usepackage[margin=2cm]{geometry}
\usepackage{comment}
\usepackage[makeroom]{cancel}
\usepackage[dvipsnames]{xcolor}
\usepackage{mathrsfs}
\usepackage{float}

\usepackage{stackengine}
\stackMath
\newcommand\tenq[2][1]{%
 \def\useanchorwidth{T}%
  \ifnum#1>1%
    \stackon[0pt]{\tenq[\numexpr#1-1\relax]{#2}}{\scriptscriptstyle\sim}%
  \else%
    \stackon[1pt]{#2}{\scriptscriptstyle\sim}%
  \fi%
}

\newcommand{\de}{\mbox{d}}
\newcommand{\lf}{\left}
\newcommand{\rg}{\right}
\newcommand{\be}{\begin{equation}}
\newcommand{\ee}{\end{equation}}
\newcommand{\pha}{\phantom{a}}

\newcommand{\pa}{\partial}

\newcommand{\bea}{\begin{eqnarray}}
\newcommand{\eea}{\end{eqnarray}}
\newcommand{\scr}{\scriptscriptstyle}

\numberwithin{equation}{section}

\begin{document}

\title{\Large Backreaction of scalar waves on black holes at low frequencies} 

\author{Marco de Cesare}
\email{marco.decesare@na.infn.it}
\affiliation{Dipartimento di Fisica ``Ettore Pancini'', Universit{\`a} di Napoli Federico II, Napoli, Italy}
\affiliation{INFN, Sezione di Napoli, Italy}

\author{Roberto Oliveri}
\email{roberto.oliveri@obspm.fr}
\affiliation{LUTH, Laboratoire Univers et Th\'eories, Observatoire de Paris, CNRS, Universit\'e PSL, Universit\'e Paris Cit\'e, 5 place Jules Janssen, 92190 Meudon, France}

\begin{abstract}
We study the accretion of a Schwarzschild black hole due to spherically symmetric perturbations sourced by a minimally coupled massless scalar field. The backreaction of the black hole to low-frequency ingoing scalar waves is computed analytically as a second-order perturbative effect, using matched asymptotic expansions to relate the behaviour of the scalar field in the vicinity of the horizon and at null infinity.
As an application of our results, we compute the mass increase due to (i) ingoing wave packets with an arbitrary profile and (ii) incoherent radiation.
Our results could serve as a model for the backreaction of environmental scalar fields on black holes.

\end{abstract}

\maketitle

\nopagebreak

%%%%%%%%%%%%%%%%%%%%%%%%%%%%%%%%%%%%%
%%%%%%%%%%%%%%%%%%%%%%%%%%%%%%%%%%%%%
\section{Introduction}

Astrophysical black holes do not exist in a vacuum. They form as a result of the gravitational collapse of matter, and their accretion dynamics after formation is determined by the non-linear interaction of gravity with matter fields. The mass of a black hole grows in time due to the flux of matter energy-momentum and gravitational waves through the black-hole horizon. In dynamical situations, different quasi-local definitions have been proposed for the black-hole horizon, among which the best known are the {\it future outer trapping horizon} \cite{Hayward:1993wb}, the {\it dynamical horizon} \cite{Ashtekar:2002ag}, or the {\it slowly evolving horizon} \cite{Booth:2003ji} (see also Refs.~\cite{Ashtekar:2004cn,Booth:2005qc,Gourgoulhon:2008pu}). These definitions replace the standard notion of the {\it event horizon}, which is teleological in nature and thus cannot be probed by any physical means.

The laws that determine the evolution of the black-hole horizon are quasi-local. However, in many practical situations one may want to recover a global picture of the spacetime, and thus relate the evolution of a black hole to the asymptotic behaviour of matter fields far from the black hole. In particular, in the case of scattering from a black hole, it is of interest to compute the resulting mass increase due to the partial absorption of waves with a given profile.
While the backreaction due to steady-state accretion of matter onto a black hole has been studied in Ref.~\cite{Babichev:2012sg} (see also Refs.~\cite{Kimura:2021dsa,Nakamura:2021mfv}), in more general dynamical situations the problem remains unexplored.
In this work, we study analytically the evolution of the black-hole mass in response to low-frequency massless scalar field perturbations originating from past null infinity in spherical symmetry. Due to the presence of the black hole, the scalar field feels an effective potential barrier. As a consequence, ingoing waves get scattered off the barrier and thus are partly absorbed by the black hole and partly reflected toward future null infinity~\cite{Futterman:1988ni}. The low frequency of the perturbations ensures that the quasi-normal modes of the system are not excited during the scattering process; hence, the reflected signal is almost undistorted. The flux of energy through the black-hole horizon results in a mass increase of the black hole itself or, equivalently, in the expansion of the future outer trapping horizon.

We adopt a perturbative approach to analyze the dynamics of the system, assuming a Schwarzschild geometry for the background and spherically symmetric perturbations for both matter and the geometry. The assumption of spherical symmetry is not too restrictive, since the low-frequency scattering cross section for a scalar field is dominated by the monopole contribution \cite{Futterman:1988ni}.
First, we solve the dynamics of the scalar field in Fourier space and in the low-frequency regime, deriving approximate analytical solutions that are valid in the regions far from the black hole and close to its Schwarzschild radius. The two asymptotics are then suitably matched in their overlap region to compute the transmission and reflection coefficients, following the approach in Ref.~\cite{Starobinsky:1973aij}. After transforming back to position space to build wave packets, we find a novel relation between the transmitted and ingoing wave packets. Next, the backreaction of the scalar wave on the black-hole mass is obtained as a second-order effect in perturbation theory. We derive for the first time a simple closed-form expression for the accretion rate in terms of the ingoing wave packet that can have arbitrary profile. The mass increase in turn leads to a decrease of the surface gravity, whose evolution is also obtained analytically.
Our results shed light on the evolution of the trapping horizon due to the scattering of massless fields, and provide us with an analytical handle that could serve as a complement to numerical relativity simulations.

We then apply our results to the special cases of coherent and incoherent radiation, computing the relative mass increase as a function of the ingoing null coordinate. In the coherent case, the mass has two plateaux: one in the past corresponding to the background value of the mass, and one in the future corresponding to the final value of the black-hole mass after the wave packet has scattered. 
Moreover, we observe two general features: i) the mass increase displays horizontal inflection points in correspondence with the inflection points of the ingoing wave packet and ii) for fixed ingoing flux, the magnitude of the mass increase is greater for more sharply peaked wave packets. This is in agreement with features observed in numerical simulations \cite{Guzman:2012jc}.
In the case of incoherent radiation, we find that the accretion rate is proportional to the ingoing flux, with a proportionality constant that is fully determined by the 
amplitude in Fourier space and by the black-hole mass.

%%%%%%%%%%%%%%%%%%%%%%%%%%%%%%%%%%%%%
%%%%%%%%%%%%%%%%%%%%%%%%%%%%%%%%%%%%%
\section{Scalar perturbations on Schwarzschild}\label{Sec:2}

Throughout this work, we focus on spherically symmetric configurations for both matter and the geometry. The most general spherically symmetric metric in four spacetime dimensions reads
\begin{equation}\label{Eq:MetricFull}
	\de s^{2}  =  -A(t,r)e^{\nu(t,r)}\de t^{2} + A^{-1}(t,r)\de r^{2} + r^{2}\de\Omega^{2}~,
\end{equation}
with $A(t,r)=1-2 m(t,r)/r$, where $m(t,r)$ is the Misner-Sharp mass (i.e., the energy contained in a spherical shell with areal radius $r$). We assume a Schwarzschild background with constant mass $M$.
We introduce matter perturbations in the form of a minimally coupled scalar field $\phi$, with energy-momentum tensor\footnote{It has been proved in Ref.~\cite{Faraoni:2021zin} that in the presence of a massless scalar field acting as null dust the metric cannot have the Vaidya form.} $T_{ab}=\pa_a\phi \pa_b\phi-\frac{1}{2}g_{ab} \lf(g^{cd}\pa_c\phi \pa_d\phi\rg)$. The background value of the scalar field $\phi$ is zero; hence, we treat $\phi$ as a first-order quantity in the perturbative expansion. Moreover, since $T_{ab}$ is quadratic in $\phi$, the presence of the scalar field only affects the geometry starting from second order in perturbation theory. The metric~\eqref{Eq:MetricFull} is then expanded perturbatively around the Schwarzschild background, with the ans{\"a}tze $m(t,r)=M+m^{(1)}(t,r)+m^{(2)}(t,r)+\dots$ and $\nu(t,r)=\nu^{(1)}(t,r)+\nu^{(2)}(t,r)+\dots$~.

Expanding the Einstein field equations $G^{a}_{\pha b}=8\pi\, T^{a}_{\pha b}$~, we find to first order in perturbation theory $\de m^{(1)}=\de \nu^{(1)}=0$. This implies that the first-order corrections $m^{(1)}$ and $\nu^{(1)}$ are constant and therefore merely amount to a redefinition of the background mass and time gauge; hence, we can set them to zero without loss of generality. Furthermore, the scalar field obeys the wave equation $\Box\phi=0$.
The dynamics of $\phi$ will be analyzed in detail in Section~\ref{Sec:3}.

To second order in the perturbative expansion, the $(t,t)$ and $(t,r)$ components of the field equations read 
\begin{subequations}\label{Eq:SecondOrderEqs}
\begin{align}
\frac{\pa m^{(2)}}{\pa r}&=2\pi r^2 \lf[\lf(1-\frac{2M}{r}\rg)^{-1}\lf(\frac{\pa\phi}{\pa t}\rg)^2+\lf(1-\frac{2M}{r}\rg)\lf(\frac{\pa\phi}{\pa r}\rg)^2 \rg]~,\label{Eq:SecondOrderEqs_tt}\\
\frac{\pa m^{(2)}}{\pa t}&=4\pi r^2 \lf(1-\frac{2M}{r}\rg)\frac{\pa\phi}{\pa t}\frac{\pa\phi}{\pa r}  ~.\label{Eq:SecondOrderEqs_tr}
\end{align}
\end{subequations}
From the $(r,r)$ component of the field equations one finds the equation for $\nu^{(2)}$; however, for our purposes we will only be concerned with $m^{(2)}$.

Introducing the Regge-Wheeler tortoise coordinate $r_* = r + 2M\log|r/(2M)-1|$ and taking the near-horizon limit $r_*\to-\infty$ of Eqs.~\eqref{Eq:SecondOrderEqs_tt},~\eqref{Eq:SecondOrderEqs_tr}, we obtain
\begin{subequations}\label{Eq:SecondOrderEqsRW}
\begin{align}
\frac{\pa m^{(2)}}{\pa r_*}&\approx 2\pi \lf[ \lf(\frac{\pa (r\phi)}{\pa t}\rg)^2+\lf(\frac{\pa (r\phi)}{\pa r_*}\rg)^2 \rg] ~,\\
\frac{\pa m^{(2)}}{\pa t}&\approx4\pi \frac{\pa (r\phi)}{\pa t}\frac{\pa (r\phi)}{\pa r_*}~.
\end{align}
\end{subequations}
Finally, introducing the null coordinates $u=t-r_*$~, $v=t+r_*$ and combining Eq.~\eqref{Eq:SecondOrderEqsRW}, we obtain 
\be\label{Eq:BackReaction_m2}
\frac{\pa m^{(2)}}{\pa v}\approx 4\pi \lf(\frac{\pa (r\phi)}{\pa v} \rg)^2~,\quad \frac{\pa m^{(2)}}{\pa u}\approx -4\pi \lf(\frac{\pa (r\phi)}{\pa u} \rg)^2~.
\ee
Equation~\eqref{Eq:BackReaction_m2} will be used in Section~\ref{Sec:4} to determine the corrections to the black-hole mass. 

%

%%%%%%%%%%%%%%%%%%%%%%%%%%%%%%%%%%%%%
%%%%%%%%%%%%%%%%%%%%%%%%%%%%%%%%%%%%%
\section{Scalar field dynamics: scattering of wave packets}\label{Sec:3}

The scalar field $\phi$ obeys the wave equation $\Box\phi=0$, which in Schwarzschild coordinates reads
\be\label{Eq:KleinGordon_r}
\frac{\pa^2 \phi}{\pa t^2} - \frac{\Delta}{r^4} \frac{\pa}{\pa r}\lf(\Delta \frac{\pa\phi}{\pa r}\rg)=0~,
\ee
with $\Delta\equiv r^2-2M r$~. Taking the Fourier transform $\phi(t,r)= (2\pi)^{-1}\int \de\omega\, R(\omega, r)  e^{-i \omega t} $ and introducing a new radial coordinate $x\equiv r/(2M)-1$, we bring Eq.~\eqref{Eq:KleinGordon_r} to the form \cite{Starobinsky:1973aij}
\be\label{Eq:KleinGordon_x}
x^2(1+x)^2\frac{\de^2 R}{\de x^2}+x(1+x)(1+2x)\frac{\de R}{\de x}+(2M\omega)^2(1+x)^4R=0~.
\ee

We focus on modes that are purely ingoing at the horizon, i.e.~$R\sim {\cal T}(\omega) e^{-i\omega r_{*}}/r$ as $r_{*}\to -\infty$, where
${\cal T}(\omega)$ is the transmission coefficient. Far from the black hole, we require that the solution be a linear superposition of ingoing and outgoing waves, $R\sim {\cal I}(\omega) e^{-i\omega r_{*}}/r+{\cal R}(\omega)e^{i\omega r_{*}}/r $ as $r_{*}\to \infty$, where ${\cal I}(\omega)$ and ${\cal R}(\omega)$ are the incidence and reflection coefficients.
In terms of the $x$ coordinate, these boundary conditions read as: $R\sim {\cal T}(\omega) x^{-i (2M\omega)}/(2M)$ as $x\to0$, and $R\sim {\cal I}(\omega) e^{-i(2M\omega) x}/(2M x)+{\cal R}(\omega)e^{i(2M\omega) x}/(2M x) $ as $x\to+\infty$. In the following, we will often omit the $\omega$ dependence to make the notation lighter.

We are interested in solving Eq.~\eqref{Eq:KleinGordon_x} in the low-frequency limit, so as not to excite the quasi-normal modes. To this end, we introduce a small dimensionless parameter $\epsilon\equiv 2M \omega$ such that $|\epsilon|\ll1$. Approximate analytical solutions for Eq.~\eqref{Eq:KleinGordon_x} can then be found in the near region, close to the Schwarzschild radius of the background geometry, and in the far region at large distances from the black hole. Next, by matching the two asymptotics, we can determine the relations between the transmission, reflection, and incidence coefficients, thus obtaining a global approximation to the solution of Eq.~\eqref{Eq:KleinGordon_x}. In the derivation of the asymptotics we follow similar steps as in Ref.~\cite{Starobinsky:1973aij,Brito:2015oca}. However, there are important subtleties in the case of monopole perturbations here considered that have not been discussed in these references.

\medskip
{\it Solution in the near region.} Here we focus on the regime $x |\epsilon| \ll 1$, where Eq.~\eqref{Eq:KleinGordon_x} can be approximated as
\be\label{Eq:KleinGordon_x_Near}
x^2(1+x)^2\frac{\de^2 R}{\de x^2}+x(1+x)(1+2x)\frac{\de R}{\de x}+\epsilon^2(1+4x)R=0~.
\ee
Note that we included a term $\sim\epsilon^2x$ in the prefactor of $R$ that arises from the linearization of $(1+x)^4$, which has been overlooked in previous works; see, e.g.,~Refs.~\cite{Starobinsky:1973aij,Brito:2015oca,Vicente:2022ivh} for $l=0$.
Equation~\eqref{Eq:KleinGordon_x_Near} can be solved exactly. With our choice of boundary conditions, the solution reads
\be\label{Eq:NearSol}
R_{\rm near}(x)=a\, x^{-i\epsilon}(1+x)^{\sqrt{3}\epsilon} ~_2F_1 \lf(1+(\sqrt{3}-i)\epsilon,(\sqrt{3}-i)\epsilon;1-2i\epsilon;-x\rg)~,
\ee
where $a$ is an integration constant and $_2F_1$ is the hypergeometric function. For small values of $x$
\be\label{Eq:NearSol_smallx}
R_{\rm near}(x) \approx a\, x^{-i\epsilon}~, \quad x\to0~,
\ee
and thus we identify the transmission coefficient to be ${\cal T}=2M a$.
In the large-$x$ limit, $R_{\rm near}$ should be matched to the far-region solution.
For this reason, we compute the leading-order asymptotics of the solution \eqref{Eq:NearSol} in the limit where $x\gg1$, retaining terms up to first order in $\epsilon$, which gives~\footnote{We note that the large-$x$ asymptotics of the hypergeometric function (used for instance in Refs.~\cite{Starobinsky:1973aij,Brito:2015oca}, where non-zero values are assumed for the black-hole spin and the angular momentum of the scalar field)
$$
~_2F_1\lf(a,b;c;-x\rg)\approx \frac{\Gamma(b-a)\Gamma(c)}{\Gamma(b)\Gamma(c-a)} x^{-a}+\frac{\Gamma(a-b)\Gamma(c)}{\Gamma(a)\Gamma(c-b)} x^{-b}
$$
does not apply in our case, since $a-b$ is an integer; see Ref.~\cite{Bateman:100233}. In this case extra care is needed. To obtain the asymptotics~\eqref{Eq:NearSol_largex}, we use the integral representation $_2F_1\lf(a,b;c;-x\rg)=\frac{\Gamma(c)}{\Gamma(b)\Gamma(c-b)}\int_0^{\infty}\de t\, t^{-b+c-1}(t+1)^{a-c}(t-x+1)^{-a}$, with $\Re(c)>\Re(b)>0$, and expand the integrand in the limit of large $x$; then, we substitute the asymptotics thus obtained in \eqref{Eq:NearSol} and expand the result to first order in $\epsilon$.}
\be\label{Eq:NearSol_largex}
R_{\rm near}(x) \approx \frac{\mathcal{T}}{2M}\lf(1+\frac{i\epsilon}{x}\rg), \quad x\to+\infty~.
\ee

\medskip
{\it Solution in the far region.} At large distances from the hole, $x\gg1$, we can approximate Eq.~\eqref{Eq:KleinGordon_x} as
\be\label{Eq:KleinGordon_x_Far}
\frac{\de^2 R}{\de x^2}+\frac{2}{x}\frac{\de R}{\de x}+\epsilon^2\lf(1+\frac{2}{x}\rg)R=0~.
\ee
The general solution of Eq.~\eqref{Eq:KleinGordon_x_Far} is a linear combination of the confluent hypergeometric function $U$ and the Kummer confluent hypergeometric function $_1F_1$
\be\label{Eq:FarSol}
R_{\rm far}(x)=e^{-i |\epsilon| x} \Big(c_1 \;U(1+i|\epsilon|,2,2i|\epsilon| x)+c_2 \;_1F_1(1+i|\epsilon|;2;2i|\epsilon| x) \Big)~,
\ee
with $c_{1}$, $c_2$ integration constants. Taking the large-$x$ limit of the solution \eqref{Eq:FarSol}, we obtain
\be\label{Eq:FarSol_largex}
R_{\rm far}(x)\approx \frac{1}{2i|\epsilon| x}\lf( (c_1-c_2) e^{-i|\epsilon| x}+ c_2e^{i|\epsilon| x} \rg), \quad x\to+\infty~.
\ee
Recalling the definition of the incidence and reflection coefficients, we find from Eq.~\eqref{Eq:FarSol_largex} that ${\cal I}=(c_1-c_2)/(2i\omega)$, ${\cal R}=c_2/(2i\omega)$ for $\omega>0$, whereas ${\cal I}=c_2/(2i|\omega|)$, ${\cal R}=(c_1-c_2)/(2i|\omega|)$ for $\omega<0$.

In the limit $x|\epsilon|\ll1$, we have
\be\label{Eq:FarSol_smallx}
R_{\rm far}(x)\approx \frac{1}{2M} \left(i \epsilon \left(\mathcal{R}-\mathcal{I}\right) + \frac{\left(\mathcal{R}+\mathcal{I}\right)\left(1+i \gamma |\epsilon|\right)}{x}\right), \quad x|\epsilon|\to0~,
\ee
where $\gamma$ is the Euler-Mascheroni's constant.

\begin{figure}
\includegraphics[width=\columnwidth]{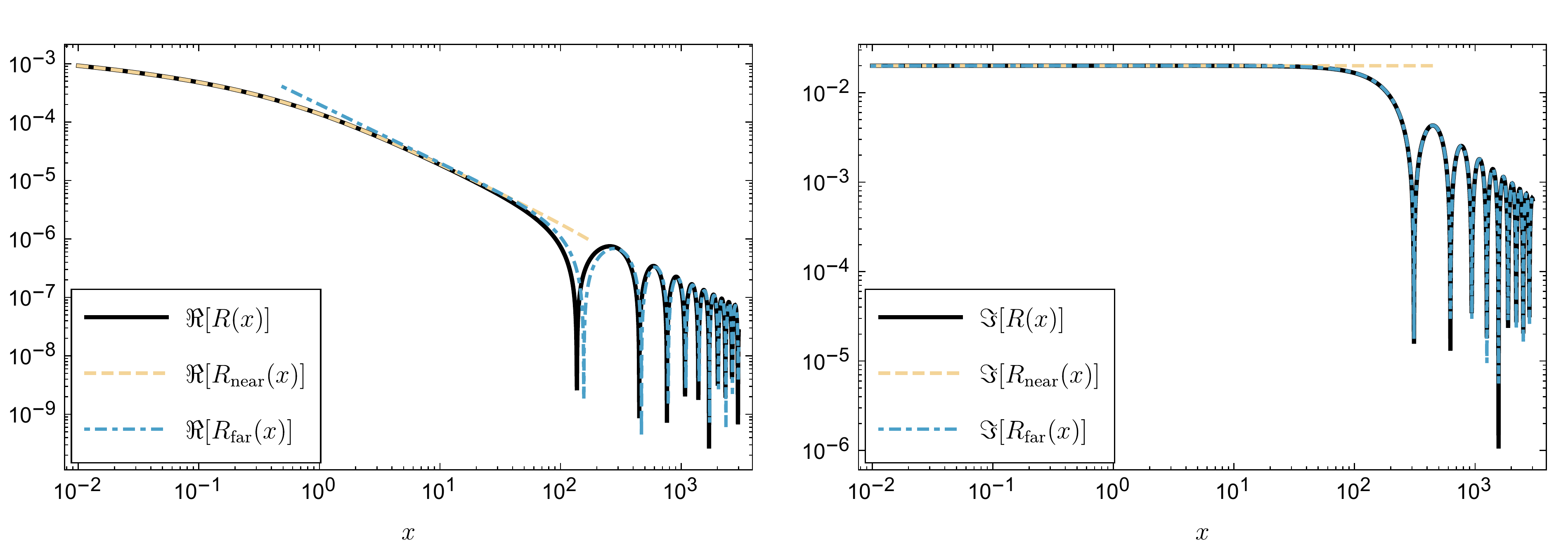}
\caption{The plots show the real (left panel) and imaginary (right panel) parts of the numerical solution of Eq.~\eqref{Eq:KleinGordon_x} satisfying ingoing boundary conditions at the horizon, along with the corresponding asymptotics \eqref{Eq:NearSol}, \eqref{Eq:FarSol} in the near and far regions, for $\epsilon=10^{-2}$ and ${\cal I}=1$, in units such that $2M=1$.
The two asymptotics are matched in their overlap region using Eq.~\eqref{Eq:CoeffTransfRefl}.}
\label{Fig:Matching}
\end{figure}

\medskip
{\it Matched asymptotics.} We can match the large-$x$ asymptotics of the near-region solution~\eqref{Eq:NearSol_largex} with the small-$x$ asymptotics of the far-region solution~\eqref{Eq:FarSol_smallx}, obtaining for the reflection and transmission coefficients\footnote{The error terms in Eq.~\eqref{Eq:CoeffTransfRefl} have been estimated by including higher-order terms in the near- and far-region asymptotics.}
\be\label{Eq:CoeffTransfRefl}
{\cal T}=-2i\epsilon\,{\cal I}+{\cal O}(\epsilon^3) ~\quad {\cal R}=\lf(-1+2\epsilon^2\rg){\cal I}+{\cal O}(\epsilon^3)~.
\ee
Thus, the relation $|{\cal T}|^2+|{\cal R}|^2= |{\cal I}|^2$ is satisfied up to ${\cal O}(\epsilon^3)$ terms.
In Fig.~\ref{Fig:Matching} we display the asymptotics corresponding to the near and far regions, matched using Eq.~\eqref{Eq:CoeffTransfRefl}, alongside the numerical solution of Eq.~\eqref{Eq:KleinGordon_x}.

\medskip
{\it Scattering of wave packets.}
Assuming an incident wave packet $\phi_{\rm in}(v,r)=(2\pi r)^{-1}\int \de\omega\, {\cal I}(\omega) e^{-i \omega v}$, the reflected and transmitted wave packets are, respectively, given by $\phi_{\rm out}(u,r)=(2\pi r)^{-1}\int \de\omega\, {\cal R}(\omega) e^{-i \omega u}$, $\phi_{\rm {\scriptscriptstyle H},in}(v,r)=(2\pi r)^{-1}\int \de\omega\, {\cal T}(\omega) e^{-i \omega v}$ (where $u\equiv t-r_*$, $v\equiv t+r_*$ are null coordinates). Reality of $\phi_{\rm in}$ implies the condition $\overline{{\cal I}}(\omega)={\cal I}(-\omega)$; furthermore, Eq.~\eqref{Eq:CoeffTransfRefl} ensures that similar conditions are also obeyed by ${\cal T}$ and ${\cal R}$. From the above, under the assumption that $\mathcal{I}(\omega)$ has support on low frequencies, we obtain for the transmitted wave packet 
\be\label{Eq:Matching_Transmitted}
\phi_{\rm {\scriptscriptstyle H},in}(v,r)= \frac{1}{2\pi r}\int_{-\infty}^{+\infty} \de \omega\, (-2i)(2M\omega){\cal I}(\omega)e^{-i\omega v}=\frac{4M}{r}\frac{\de}{\de v}\lf(r\phi_{\rm in}(v,r)\rg)~,
\ee
while the reflected packet reads\footnote{The total derivatives in the last steps of Eqs.~\eqref{Eq:Matching_Transmitted} and \eqref{Eq:Matching_Reflected} are due to the fact that the quantity $r\phi_{\rm in}(v,r)$ has no dependence on $r$.}
\be\label{Eq:Matching_Reflected}
\phi_{\rm out}(u,r)= \frac{1}{2\pi r}\int_{-\infty}^{+\infty} \de\omega \lf(-1+2(2M\omega)^2\rg){\cal I}(\omega)e^{-i\omega u}= -\phi_{\rm in}(u,r) - \frac{8M^2}{r}\frac{\de^2}{\de u^2}\lf(r\phi_{\rm in}(u,r)\rg)~.
\ee
The ingoing and outgoing energy fluxes are, respectively,
\begin{equation}
{\cal F}_{\rm in}(v)=4\pi r^2\left(\frac{\pa}{\pa v}\phi_{\rm in}(v,r)\right)^2~,\quad  {\cal F}_{\rm out}(u)=4\pi r^2\left(\frac{\pa}{\pa u}\phi_{\rm out}(u,r)\right)^2~. 
\end{equation}
The fraction of energy absorbed by the hole is
\be\label{Eq:AbsorptionCoefficient}
Z=1-\frac{\int_{-\infty}^{+\infty}\de u\, {\cal F}_{\rm out}(u)}{\int_{-\infty}^{+\infty}\de v\, {\cal F}_{\rm in}(v)}~.
\ee

%%%%%%%%%%%%%%%%%%%%%%%%%%%%%%%%%%%%%
%%%%%%%%%%%%%%%%%%%%%%%%%%%%%%%%%%%%%
\section{Backreaction effects on the black-hole mass}\label{Sec:4}

\begin{figure}
\begin{center}
\includegraphics[scale=0.35]{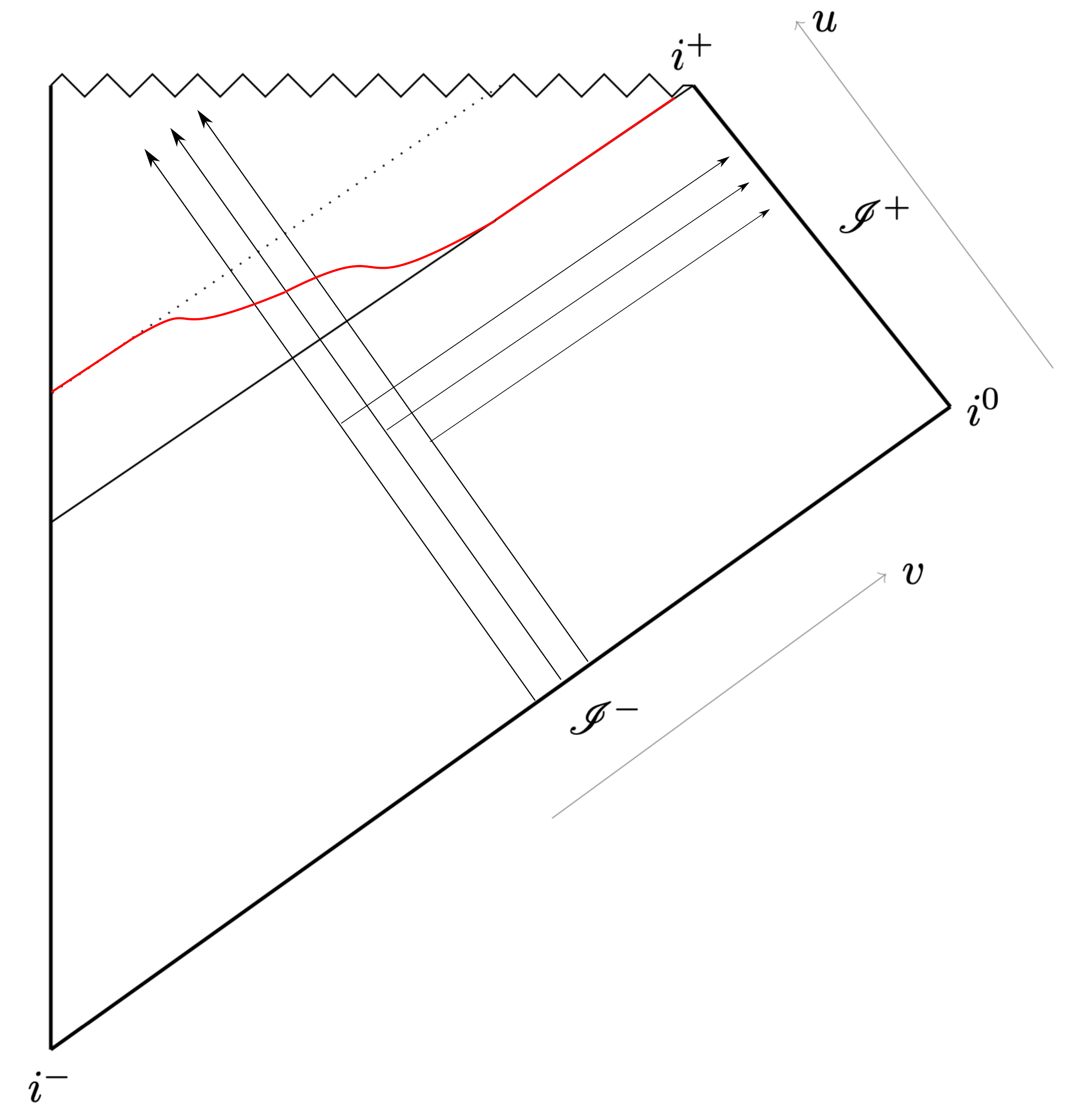}
\end{center}
\caption{Carter-Penrose diagram depicting the scattering of a wave packet by a spherically symmetric black hole and the consequent backreaction on the geometry. The {\it future outer trapping horizon} (red curve) is spacelike and interpolates between the event horizon of the Schwarzschild background (dotted line) and the event horizon of the perturbed spacetime (thick line).}
\label{fig:conformal_diagram}
\end{figure}

The backreaction effects of matter on the black-hole mass are determined by Eq.~\eqref{Eq:BackReaction_m2}. We note that in the general case the scalar field may include both an ingoing and an outgoing component in the near-horizon region.
Assuming ingoing boundary conditions as in Section~\ref{Sec:3}, Eq.~\eqref{Eq:BackReaction_m2} gives $m^{(2)}\approx m^{(2)}(v)$ in the $u\to+\infty$ limit where the horizon of the background is approached. This implies that the mass of the perturbed black hole is $\overline{M}(v)=M+\Delta M(v)$, where $\Delta M(v)\equiv \lim_{u\to+\infty}m^{(2)}(u,v)$.  Therefore, combining Eqs.~\eqref{Eq:BackReaction_m2} and  \eqref{Eq:Matching_Transmitted}, we obtain our main result
\be\label{Eq:BackReaction_matched}
\frac{\de (\Delta M)}{\de v}\approx 
64\pi M^2 \lf(\frac{\de^2\lf(r \phi_{\rm in}\rg)}{\de v^2} \rg)^2~.
\ee
Equation~\eqref{Eq:BackReaction_matched} also determines the black-hole trapping horizon as $r_{H}(v)=2(M+\Delta M(v))$.~\footnote{More precisely, this is a future outer trapping horizon according to Hayward's definition \cite{Hayward:1993wb}, since $\theta_{l}=0$, $\theta_n<0$, and $\pounds_{n}\theta_l<0$ (where $l^a$ and $n^a$ are future pointing radial null vectors, respectively outward and inward directed). In the case at hand, the last condition follows from the fact that the scalar field obeys the null energy condition, see also Refs.~\cite{deCesare:2022aoe,Booth:2005ng}.}
Integrating Eq.~\eqref{Eq:BackReaction_matched}, we obtain the mass of the evolving black hole
\be\label{Eq:FinalMass}
\overline{M}(v)=M+64\pi M^2\int_{-\infty}^{v}\de v\, \lf(\frac{\de^2\lf(r \phi_{\rm in}\rg)}{\de v^2} \rg)^2~.
\ee
Combining Eq.~\eqref{Eq:FinalMass} and Eq.~\eqref{Eq:AbsorptionCoefficient}, we find the absorption coefficient $Z\approx(M_{\rm final}-M)/{\cal F}_{\rm in}^{\rm\scr TOT}$, where $M_{\rm final}  \equiv \lim_{v \to +\infty}\overline{M}(v)$ and ${\cal F}_{\rm in}^{\rm\scr TOT}\equiv\int_{-\infty}^{\infty} \de v\, {\cal F}_{\rm in}(v)$ is the total ingoing energy flux. A Carter-Penrose diagram representing the evolution of a black hole in response to the scattering of a wave packet is shown in Fig.~\ref{fig:conformal_diagram}. 

A geometric definition for the surface gravity of a dynamical black hole was given in Ref.~\cite{Hayward:2008jq}, where it reads as $\kappa\equiv \frac{1}{2} (*d*dr)$,~\footnote{With $*$ and $d$ we denote, respectively, the Hodge dual and the exterior derivative in the $(t,r)$ space normal to the two-sphere.} which in spherical symmetry and in coordinates $(v,r)$ reduces to $\kappa(v)=\lf(1-2\frac{\pa m (v,r)}{\pa r}\rg)/(4\overline{M}(v))$~, where the rhs is evaluated at the horizon.
Changing coordinates and using Eq.~\eqref{Eq:SecondOrderEqsRW}, we obtain $\frac{\partial m(v,r)}{\pa r}\propto \lf(\frac{\pa(r\phi)}{\pa u}\rg)^2$ in the $r_*\to-\infty$ limit, which vanishes for purely ingoing boundary conditions at the horizon. Hence, we obtain 
\begin{equation}
\kappa(v)=\frac{1}{4\overline{M}(v)}~,
\end{equation}
which is monotonically decreasing.
For any given profile of the scalar field, including the cases considered in the next Section~\ref{Sec:5}, $\kappa(v)$ can be computed analytically.

%%%%%%%%%%%%%%%%%%%%%%%%%%%%%%%%%%%%%
%%%%%%%%%%%%%%%%%%%%%%%%%%%%%%%%%%%%%
\section{Physical applications}\label{Sec:5} 

Equation~\eqref{Eq:BackReaction_matched} predicts that the black-hole mass is monotonically increasing in time, consistently with Hayward's {\it area increase theorem} \cite{Hayward:1993wb}, since the scalar field obeys the null energy condition. Moreover, Eq.~\eqref{Eq:BackReaction_matched} gives the exact accretion rate of the black-hole mass as a response to the infalling scalar field, given an arbitrary profile for the ingoing wave $\phi_{\rm in}$ (so long as its power spectrum is peaked on low frequencies). It applies for coherent as well as incoherent radiation, which we analyze in the following.

\begin{figure}
\includegraphics[width=\columnwidth]{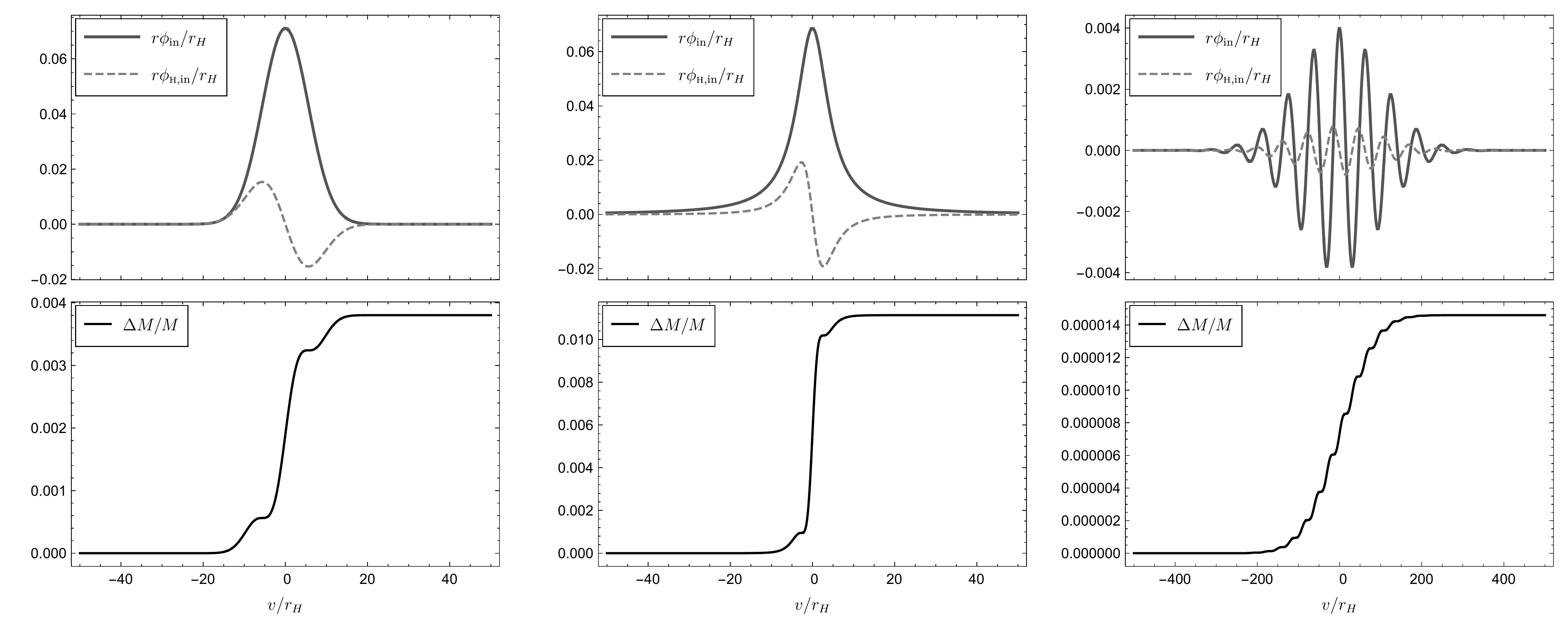}
\caption{We show the evolution of the black-hole mass corresponding to different profiles for the ingoing pulse. The leftmost and central figures correspond to a Gaussian and Lorentzian wave packets, respectively, with the same normalization and total energy flux. More specifically, these are given by $\phi_{\rm in}(v,r)=\frac{A}{\sqrt{2\pi}\sigma r}\exp\lf(-\frac{(v-v_0)^2}{2\sigma^2}\rg)$ and $\phi_{\rm in}(v,r)=\frac{A}{\pi r}\frac{\gamma}{\gamma^2+(v-v_0)^2}$, with $A=r_{H}^2$, ${\cal F}_{\rm in}^{\rm\scriptscriptstyle TOT}=10^{-2}\,r_H$ (the values of $\sigma$ and $\gamma$ are fixed accordingly: $\sigma\simeq 5.617 \,r_H$, $\gamma\simeq 4.642\, r_H$), where $r_H$ is the background Schwarzschild radius. 
The origin of the horizontal axis has been shifted so that $v_0=0$. The rightmost plots correspond instead to a modulated Gaussian $\phi_{\rm in}(v,r)=\frac{A}{\sqrt{2\pi}\sigma r}\exp\lf(-\frac{(v-v_0)^2}{2\sigma^2}\rg)\cos\lf(\omega(v-v_0) \rg)$, with $A=r_{H}^2$, $\omega=10^{-1}/r_{H}$ and $\sigma=10^2\, r_{H}$. We note the multiple horizontal inflection points corresponding to the inflection points of the ingoing pulse.}
\label{Fig:Accretion}
\end{figure}

\medskip
{\it Coherent radiation.} 
As an example, we compute the relative mass increase due to infalling wave packets with different profiles, as shown in Fig.~\ref{Fig:Accretion}.  As a general feature, for a fixed ingoing flux, the final mass is higher for wave packets that are more sharply peaked. Moreover, inflection points of the wave packets correspond to the horizontal inflection points of the evolving black-hole mass. In turn, these correspond to inflection points of the monotonically decreasing surface gravity.

\medskip
{\it Incoherent radiation.} We consider ingoing radiation with ${\cal I}(\omega)=A(\omega) e^{-i\varphi_{\omega}}$, with a random phase $\varphi_{\omega}$. This is in part similar to the {\it random phase model} for wave dark matter \cite{Hui:2021tkt}, although in our case the scalar field is massless (whereas wave dark matter must be non-relativistic). The amplitude $A(\omega)$ is real and needs not be stochastic. The two-point function is assumed to be $\langle e^{- i\varphi_{\omega}} e^{i\varphi_{\omega^{\prime}}}\rangle=2\pi C\, \delta(\omega-\omega^{\prime})$, where $C$ is a positive constant with dimensions of frequency. 
The expectation value of the ingoing flux is
\be
\langle{\cal F}_{\rm in}(v)\rangle=2C \int_{-\infty}^{+\infty}\de\omega\, A(\omega)A(-\omega) \omega^2~.
\ee
Note that $\langle{\cal F}_{\rm in}(v)\rangle$ does not depend on $v$ in this model.
Then, taking the expectation value of Eq.~\eqref{Eq:BackReaction_matched} we obtain
\be
\bigg\langle\frac{\de (\Delta M)}{\de v}\bigg\rangle \approx 32C M^2 \int_{-\infty}^{+\infty}\de\omega\, A(\omega)A(-\omega)\omega^4  \approx 16  \left(\frac{\int_{-\infty}^{+\infty}\de\omega\, A(\omega)A(-\omega)\omega^4}{\int_{-\infty}^{+\infty}\de\omega\, A(\omega)A(-\omega) \omega^2}\right)M^2\langle{\cal F}_{\rm in}\rangle ~.
\ee
Thus, the accretion rate is fully determined by the second and fourth moments of the product $A(\omega)A(-\omega)$.
In particular, if we assume a Gaussian profile $A(\omega)={\cal A} \exp(-(\omega-\bar{\omega})^2/(2\sigma^2) )$, we obtain\footnote{Note that, within our low-frequency approximation, $A(\omega)$ must be peaked on low frequencies, i.e.~$\bar{\omega}M\ll1$ and $\sigma M\ll1$.}
\be
\bigg\langle\frac{\de (\Delta M)}{\de v}\bigg\rangle \approx 24 \sigma^2 M^2 \langle{\cal F}_{\rm in}\rangle ~.
\ee
Once again we observe that the accretion rate depends on the variance of the incoming wave packets and it is greater for larger $\sigma$. The expectation value of the surface gravity $\langle \kappa(v)\rangle$ is monotonically decreasing with a constant rate in this model.

%%%%%%%%%%%%%%%%%%%%%%%%%%%%%%%%%%%%%
%%%%%%%%%%%%%%%%%%%%%%%%%%%%%%%%%%%%%
\section{Discussion}\label{Sec:6}

We investigated the low-frequency response of a Schwarzschild black hole to wave packets emitted in the far past. Our approach relies on second-order perturbation theory for the Einstein field equations with a minimally coupled massless scalar field obeying purely ingoing boundary conditions at the horizon.
Our main results are presented in Section~\ref{Sec:4} and shed light into the evolution of the trapping horizon as a result of the backreaction of low-frequency scalar waves.
Equation~\eqref{Eq:BackReaction_matched} gives a closed-form formula for the accretion rate of the black hole that is controlled by the profile of the ingoing wave.
The trapping horizon can be readily obtained from the evolution of the mass, Eq.~\eqref{Eq:FinalMass}. For wave packets, the trapping horizon interpolates between the event horizon of the Schwarzschild background and that of the perturbed spacetime, while for incoherent radiation, the horizon grows with a constant rate.
Furthermore, our analytical results explain the qualitative features observed in numerical simulations of a black hole accreting wave packets of scalar radiation~\cite{Guzman:2012jc}.
Our results could be relevant to model environmental effects in the evolution of black holes. 
In fact, scalar fields play an important role in cosmology, where they are responsible for the accelerated expansion of the universe at early times, and may play the role of dark matter at late times \cite{Hui:2021tkt}. In future work, we will extend our analysis to non-asymptotically flat spacetimes and include scalar fields with a general potential, to study the evolution of black holes in these scenarios. Moreover, our setup could be generalized further to massive complex scalar fields, which can support a black-hole hair \cite{Herdeiro:2014goa} and may form clouds that are responsible for relativistic drag forces exerted on black holes~\cite{Vicente:2022ivh,Traykova:2023qyv}.
Our analysis can be further generalized to also include hydrodynamic matter, which would allow for a comparison of our setup with recent studies on Bondi-Hoyle-Lyttleton accretion \cite{Cruz-Osorio:2023wev}.
This work can be further extended in several directions. One can attempt to move beyond the low-frequency regime, either by matching the near and far-zone asymptotics obtained for a finite frequency $\omega$, or by including higher-order corrections in the expansion for small $\epsilon = 2M\omega$, and then solving the second-order perturbative equations to compute the backreaction on the mass.
It would also be interesting to understand how the backreaction of massless fields is affected by a non-zero angular momentum of the black hole; while there are some numerical studies on this topic~\cite{Macedo:2013afa,Leite_2016}, analytically this question remains unexplored. Lastly, the cosmological analog of the classical backreaction problem considered in this paper will help us to shed light on the  effects of cosmic expansion and more general matter fields on the evolution of primordial black holes. We will address these open questions in upcoming work.

\section*{Acknowledgments}
We thank Eric Gourgoulhon and Carlos Herdeiro for helpful comments on a previous version of the manuscript.
The work of MdC is supported by Ministero dell'Universit{\`a} e Ricerca (MUR) (Bando PRIN 2017, Codice Progetto: 20179ZF5K5\_006) and by INFN (Iniziative specifiche QUAGRAP and GeoSymQFT). The work of RO is supported by the R\'egion \^Ile-de-France within the DIM ACAV$^{+}$ project SYMONGRAV (Sym\'etries asymptotiques et ondes gravitationnelles).

\bibliography{refs_backreaction}
\end{document}